\def\aa{{A\&A}}
\def\aas{{A\&AS}}
\def\aj{{AJ}}
\def\annrev{{ARA\&A}}
\def\apj{{ApJ}}
\def\apjs{{ApJS}}
\def\baas{{BAAS}}
\def\mnras{{MNRAS}}
\def\nat{{Nature}}
\def\pasp{{PASP}}

\documentclass[cup5b]{caps}

\usepackage{graphicx}
\usepackage{amssymb}
\usepackage{ociwsymp1e}
\HeadText{P. Amaro-Seoane and R. Spurzem}

\begin{document}

\pagenumbering{arabic}

\author[]{P. AMARO-SEOANE and R. SPURZEM\\Astronomisches Rechen-Institut\\
M\"onchhofstra\ss e 12 \-- 14, Heidelberg, Germany}

\chapter{Dense Gas-Star Systems: Evolution of Supermassive Stars}

\begin{abstract}
In the 60s and 70s super-massive central objects 
(from now onwards SMOs) were thought to be the main source of active galactic
nuclei (AGNs) characteristics (luminosities 
of $L \approx 10^{12} L_{\odot}$). 
The release of gravitational binding energy by the accretion of material 
on to an SMO in the range of $10^7 - 10^9 M_{\odot}$ has been suggested 
to be the primary powerhouse (Lynden-Bell 1969).
 That rather exotic idea in early time 
 has become common sense nowadays. Not only our
 own galaxy harbours a few million-solar mass black hole 
(Genzel 2001)
 but also many of other non-active galaxies show kinematic
 and gas-dynamic evidence of these objects (Magorrian et al. 1998)
The concept of central 
super-massive stars (SMSs henceforth) (${\cal M} \ge 5 \times 10^4 M_{\odot}$,
where ${\cal M}$ is the mass of the SMS)
embedded in dense stellar systems was suggested as a possible explanation 
for high-
energy emissions phenomena occurring in AGNs and quasars (Vilkoviski 
1976, Hara 1978), such as X-ray emissions (Bahcall and Ostriker, 1975).
SMSs and super-massive black holes (SMBHs) are two possibilities 
to explain the nature of SMOs, and SMSs may be 
an intermediate step towards the formation of SMBHs (Rees 1984).
In this paper we give the equations that describe the dynamics of
such a dense star-gas system which are the basis for the code that
will be used in a prochain future to simulate this scenario. We also
briefly draw the mathematical fundamentals of the code.
\end{abstract}

\section{The gaseous model}

To go from stationary to dynamical models we use a gaseous model of
star clusters 
in its anisotropic version.
It is based on the basic assumptions that:
the system can be described by a one particle distribution function,
the secular evolution is dominated by the cumulative 
effect of small angle deflections with small impact parameters (Fokker-Planck
approximation, good for
large $N$-particle systems) and 
that the effect of the two-body relaxation can be modelled by a local
heat flux equation with an appropriately tailored conductivity.

\noindent
The first assumption justifies a kinetic equation of the Boltzmann type
with the inclusion of a collisional term of the Fokker-Planck (FP) type:

\begin{equation}
\frac{\partial f}{\partial t}+v_{r}\frac{\partial f}{\partial r}+\dot{v_{r}}
\frac{\partial f}{\partial v_{r}}+\dot{v_{\theta}}\frac{\partial f}{\partial 
v_{\theta}}+\dot{v_{\varphi}}\frac{\partial f}{\partial v_{\varphi}}=\Bigg
( \frac{\delta f}{\delta t} \Bigg)_{FP}
\end{equation}
\noindent
In spherical symmetry polar coordinates $r$, $\theta$, $\phi$
are used and $t$ denotes the time. The vector
${\bf v} = (v_i), i=r,\theta,\phi$ denotes the velocity
in a local Cartesian coordinate system at the spatial point
$r,\theta,\phi$. The distribution function $f$ is a function of,
$r$, $t$, $v_r$, $v_\theta^2+v_\phi^2$ only due to spherical symmetry.
By multiplication of the Fokker-Planck equation (1.1) with
various powers of the velocity components we get up to second order a
set of moment equations which is equivalent to gas-dynamical equations
coupled with Poisson's equation: A mass equation, a continuity equation,
an Euler equation (force), radial and tangential energy equations.
The system of equations is closed by a phenomenological heat flux equation
for the flux of radial and tangential r.m.s. kinetic energy, both in
radial direction.

\section{Interaction terms for the star component}

We now introduce the interaction terms to be added to right hand of the star component
equations. 

\subsection{Equation of continuity}
In the paper by Langbein et al. (1990) they derive the interaction terms to be
added to the basic equations of the gaseous model. According to them,
the star continuity equation is no longer
\begin{equation}
\frac {\partial \rho_{*}}{\partial {\rm t}}+ \frac{1}{{\rm r}^2} \frac{\partial}
{\partial {\rm r}}({\rm r^2} \rho_{*} {\rm u_{*}})=0,
\end{equation}
\noindent
but
\begin{equation}
\frac {\partial \rho_{*}}{\partial {\rm t}}+ \frac{1}{{\rm r}^2} \frac{\partial}
{\partial {\rm r}}({\rm r^2} \rho_{*} {\rm u_{*}})=\bigg( \frac{\delta \rho_{*}}
{\delta {\rm t}} \bigg)_{\rm coll}+\bigg( \frac{\delta \rho_{*}}
{\delta {\rm t}} \bigg)_{\rm lc};
\end{equation}
\noindent
where the right-hand term reflects the time variation of the star's density due to 
stars interactions (i.e. due to the calculation of the mean rate of gas production by 
stars collisions) and loss-cone (stars plunging onto the central object). 

\noindent
If $f(v_{\rm rel})$ is the stellar distribution of relative velocities,
then the mean rate of gas production by stellar collisions is
\begin{equation}
\bigg( \frac{\delta \rho_{*}}{\delta {\rm t}} \bigg)_{\rm coll}=
-\int_{|v_{\rm rel|}>\sigma_{\rm coll}} \frac{\rho_{*}f_{c}(v_{\rm 
rel})}{t_{\rm coll}} f(v_{\rm rel})d^3v_{\rm rel}
\end{equation}
$f(v_{\rm rel})$ is a Schwarzschild-Boltzmann distribution,
\begin{equation}
f(v_{\rm rel})=\frac{1}{2 \pi^{3/2} \sigma_{r} \sigma_{t}^2} \cdot {\rm exp}
\Bigg( -\frac{(v_{\rm rel,r}-u_{*})^2}{4 \sigma_{r}^2}-\frac{v_{\rm rel,t}^2}
{2 \sigma_{t}^2}  \Bigg)
\end{equation}

\noindent
As regards $f_{c}$, it is the relative fraction of mass liberated per stellar
collision into the gaseous medium. Under certain assumptions given in the 
initial work of Spitzer \& Saslaw (1966), we can calculate it as an average
over all impact parameters resulting in $r_{\rm min}<2r_{*}$ and as a function
of the relative velocity at infinity of the two colliding stars, $v_{\rm rel}$.
Langbein et al. (1990) approximate their result by

\[ f_{c}(v_{\rm rel}) = \left\{ \begin{array}{ll}
  \big(1+q_{\rm coll} \sqrt{\sigma_{\rm coll}/v_{\rm rel}} \big)^{-1}& 
\mbox{$v_{\rm rel} > \sigma_{\rm coll}$}\\
  0 & \mbox{$v_{\rm rel} < \sigma_{\rm coll}$},
  \end{array} 
   \right. \]
\noindent
with $q_{\rm coll}=100$. So, we have that

\[ f_{c}(v_{\rm rel}) = \left\{ \begin{array}{ll}
  0.01 & \mbox{$\sigma_{\rm coll}=v_{\rm rel}$}\\
  0 & \mbox{$\sigma_{\rm coll}>v_{\rm rel}$},
  \end{array} 
   \right. \]

\noindent
${t_{\rm coll}}$ is defined as the mean time which has passed 
when the number of stars within a volume $V=\Sigma \cdot v_{\rm rel} \cdot 
\triangle t$ is one, where $v_{\rm rel}$ is the relative velocity at 
infinity of two colliding stars. 

\noindent
Computed for an average distance of closest approach $\bar{r}_{\rm min}=
\frac{2}{3}r_{*}$, this time is
\begin{equation}
n_{*}V(t_{\rm coll})=1= n_{*}\Sigma v_{\rm rel} t_{\rm coll}.
\end{equation}
\noindent
And so,
\begin{equation}
t_{\rm coll}=\frac{m_{*}}{\rho_{*}\Sigma\sigma_{\rm rel}},
\end{equation}
\noindent
with
\begin{equation}
\Sigma=\pi \bar{r}_{\rm min}^2 \bigg( 1+ \frac{2Gm_{*}}
{\bar{r}_{\rm min}\sigma_{\rm rel}^2}\bigg);
\end{equation}
$\sigma_{\rm rel}^2=2\sigma_{*}^2$ is the stellar velocity dispersion and
$\Sigma$ a collisional cross section with gravitational focusing.

The first interaction term is \footnote{In the paper there
are two typos in the equation, the correct signs and factors are given here.}
\begin{equation}
\bigg( \frac{\delta \rho_{*}}{\delta {\rm t}} \bigg)_{\rm coll}=- \rho_{*} 
\frac{\rm f_{c}}{\rm t_{coll}} \bigg[ 1-{\rm erf} \bigg( \frac{\sigma_{\rm coll}}
{\sqrt {6}\sigma_{\rm r}} \bigg) \bigg]\bigg[ 1-{\rm erf} \bigg( \frac
{\sigma_{\rm coll}}{\sqrt {6}\sigma_{\rm t}} \bigg) \bigg]^2 
\end{equation}
\noindent
which, for simplification, we re-call like this
\begin{equation}
\bigg( \frac{\delta \rho_{*}}{\delta {\rm t}} \bigg)_{\rm coll}\equiv -\rho_{*} 
{\rm X_{coll}}.
\end{equation}
\noindent

Since the evolution of the system that we are studying can be regarded as
stationary, we introduce for each equation the ``logarithmic variables'' in 
order to study the evolution at long-term. In
the other hand, if the system happens to have quick changes, we should use
the ``non-logarithmic'' version of the equations. For this reason we will write
at the end of each subsection the equation in terms of the logarithmic variables.

In the case of the equation of continuity, we develop it and divide it by $\rho_{*}$
because we are looking for the logarithm of the stars density, $\partial \ln \rho_{*}
/\partial t=(1/\rho_{*})\partial \rho_{*}/ \partial t$. 
The result is:

\begin{equation}
\frac{\partial \ln \rho_{*}}{\partial t} + \frac{\partial u_{*}}{\partial r}+
u_{*} \frac{\partial \ln \rho_{*}}{\partial r}+ \frac{2u_{*}}{r}= \frac{1}
{\rho_{*}}\bigg( \frac{\delta \rho_{*}}{\delta {\rm t}} \bigg)_{\rm coll}+
\frac{1}{\rho_{*}}\bigg( \frac {\delta \rho_{*}} {\delta {\rm t}} \bigg)_{\rm lc}
\end{equation}

\subsection{Momentum balance}

The second equation has the following star interaction terms:

\begin{equation}
\frac{\partial u_{*}}{\partial t}+u_{*} \frac{\partial u_{*}}{\partial r} + {GM_r\over r^2} +
{1\over\rho_{*}}\frac{\partial p_r}{\partial r} + 2{p_r - p_t\over\rho_{*} r} = 
\bigg( \frac{\delta u_{*}}{\delta t}\bigg)_{\rm drag}  
\end{equation}
\noindent
The interaction term is due to the decelerating force at which stars that move
inside the gas are subject to. Explicitely, it is
\begin{equation}
\bigg( \frac{\delta u_{*}}{\delta t}\bigg)_{\rm drag} = - X_{\rm drag}\frac{1}{\rho_{*}} 
(u_{*}-u_{g})
\end{equation}
\noindent
where we have introduced the following definition:
\begin{equation}
X_{\rm drag} \equiv -C_{D} \frac{\pi r_{*}^2}{m_{*}}\rho_{*} \rho_{g} \sigma_{\rm tot},
\end{equation}
\noindent
with $\sigma_{\rm tot}^2=\sigma_{r}^2+\sigma_{t}^2+(u_{*}-u_{g})^2$
\vskip0.3cm
To the end of the calculation of the logarithmic variable version of the equation, we
multiply by $\rho_{*} r/p_{r}$:
\begin{equation}
\frac{\rho_{*} r}{p_{r}} \bigg( \frac{\partial u_{*}}{\partial t}+ u_{*} \bigg)+
\frac{GM_{r}}{rp_{r}}\rho_{*}+\frac{\partial \ln p_{r}}
{\partial \ln r}+2(1-\frac{p_{t}}{p_{r}})=-X_{\rm drag} \frac{r}{p_{r}}(u_{*}-u_{g})
\end{equation}

\subsection{Radial energy equation}

As regards the last but one equation, the interaction terms are:
\begin{equation}
\frac{\partial{p_r}}{\partial {t}} + \frac{1}{r^2} \frac{\partial}{\partial r} 
(r^2 u_{*} p_{r})+2 p_{r} \frac{\partial u_{*}}{\partial r}+\frac{4}{5} \frac{(2p_{r}-p_{t})}
{t_{\rm relax}} +  \frac{1}{r^2} \frac{\partial}{\partial r} 
(r^2 F_{r})- \frac{2F_{t}}{r}= \nonumber
\end{equation}
\begin{equation}
\bigg( \frac{\delta p_{r}}{\delta t}\bigg)_{\rm drag}+\bigg( \frac{\delta p_{r}}
{\delta t}\bigg)_{\rm coll},
\label{eqn:pr} 
\end{equation}
\noindent
where
\begin{equation}
\bigg( \frac{\delta p_{r}}{\delta t}\bigg)_{\rm drag}=-2X_{\rm drag} \sigma_{r}^2,~
\bigg( \frac{\delta p_{r}}{\delta t}\bigg)_{\rm coll}=-X_{\rm coll} 
\rho_{*} \tilde{\sigma_{r}}^2 \epsilon. 
\end{equation}
\noindent
In order to determine $\epsilon$ we introduce the ratio $k$ of kinetic energy of the
remaining mass after the encounter over its initial value (before the encounter); $k$
is a measure of the inelasticity of the collision: for $k=1$ we have the minimal inelasticity,
just the kinetic energy of the liberated mass fraction is dissipated, whereas if $k<1$
a surplus amount of stellar kinetic energy is dissipated during the collision
(tidal interactions and excitation of stellar oscillations). If we calculate the energy loss
in the stellar system per unit volume as a function of $k$ we obtain
\begin{equation}
\epsilon=f_{c}^{-1}[1-k(1-f_{c})].
\end{equation}

We divide by $p_{r }$ so that we get the logarithmic variable version of the equation.
We make also the following substitution:
\begin{eqnarray}
&F_{r}=&3p_{r}v_{r} \nonumber \\
&F_{t}=&2p_{t}v_{t}
\end{eqnarray}
\noindent
The resulting equation is
\begin{eqnarray}
\frac{\partial \ln p_{r}}{\partial t}+(u_{*}+3v_{r}) \frac{\partial \ln p_{r}}{\partial r}
+3 \bigg( \frac{\partial u_{*}}{\partial r}+\frac{\partial v_{r}}{\partial r} \bigg)+
\frac{2}{r}\bigg( u_{*}+3v_{r}-2v_{t} \frac{p_{t}}{p_{r}} \bigg)+ \nonumber \\
\frac{4}{5}\frac{ 2-\frac{p_{t}} {p_{r}}}{t_{\rm relax}}= \frac{1}{p_{r}} 
\bigg( \frac{\delta p_{r}}{\delta t}\bigg)_{\rm drag}+\frac{1}{p_{r}}
\bigg( \frac{\delta p_{r}} {\delta t}\bigg)_{\rm coll}
\end{eqnarray}

\subsection{Tangential energy equation}

To conclude the set of equations of the star component
with the interaction terms, we have the following equation:

\begin{equation}
\frac{\partial{p_t}}{\partial {t}} + \frac{1}{r^2} \frac{\partial}{\partial r} 
(r^2 u_{*} p_{t})+2 \frac{p_{t}u_{*}}{r}-\frac{4}{5} \frac{(2p_{r}-p_{t})}{t_{\rm relax}}+
 \frac{1}{r^2} \frac{\partial}{\partial r}(r^2F_{t})+\frac{2F_{t}}{r}= \nonumber
\end{equation}
\begin{equation}
\bigg( \frac{\delta p_{t}}{\delta t}\bigg)_{\rm drag}+\bigg( \frac{\delta p_{t}}
{\delta t}\bigg)_{\rm coll}, 
\end{equation}
\noindent
where
\begin{equation}
\bigg( \frac{\delta p_{t}}{\delta t}\bigg)_{\rm drag}=-2X_{\rm drag} \sigma_{t}^2,~
\bigg( \frac{\delta p_{t}}{\delta t}\bigg)_{\rm coll}=-X_{\rm coll} 
\rho_{*} \tilde{\sigma_{t}}^2 \epsilon. 
\end{equation}
\noindent

We follow the same path like in the last case and so we get the following logarithmic
variable equation:

\begin{eqnarray}
\frac{\partial \ln p_{t}}{\partial t}+(u_{*}+2v_{t}) \frac{\partial \ln p_{t}}{\partial r}
+\frac{\partial}{\partial r}(u_{*}+2v_{t})+ \frac{4}{r}(u_{*}+2v_{t})- \\
\frac{4}{5} \frac{2\frac{p_{r}}{p_{t}}-1}{t_{\rm relax}}= \frac{1}{p_{t}} 
\bigg( \frac{\delta p_{t}}{\delta t}\bigg)_{\rm drag}+
\frac{1}{p_{t}} \bigg( \frac{\delta p_{t}} {\delta t}\bigg)_{\rm coll} \nonumber
\end{eqnarray}

\section{The gaseous component and the interaction terms}

In this section we give the set of equations corresponding to the
gaseous component as for their right hand interaction terms.

\subsection{Equation of continuity}

For the SMS the equation of continuity looks as follows:

\begin{equation}
\frac {\partial \rho_{g}}{\partial { t}}+ \frac{1}{{\rm r}^2} \frac{\partial}
{\partial { r}}({\rm r^2} \rho_{g} { u_{g}})=\bigg( \frac{\delta \rho_{g}}
{\delta { t}} \bigg)_{\rm coll}
\end{equation}
\noindent
where, for the mass conservation, we have that, obviously,
\begin{equation}
\bigg( \frac{\delta \rho_{g}}{\delta { t}} \bigg)_{\rm coll}=
-\bigg( \frac{\delta \rho_{*}}{\delta { t}} \bigg)_{\rm coll}
\end{equation}

We follow the same procedure as for the star continuity equation to get the 
equation in terms of the logarithmic variables:

\begin{equation}
\frac{\partial \ln \rho_{g}}{\partial t} + \frac{\partial u_{g}}{\partial r}+
u_{g} \frac{\partial \ln \rho_{g}}{\partial r}+ \frac{2u_{g}}{r}= \frac{1}
{\rho_{g}}\bigg( \frac{\delta \rho_{g}}{\delta {\rm t}} \bigg)_{\rm coll}
\end{equation}

\noindent
The interaction term is in this case

\begin{equation}
\frac{1}{\rho_{g}}\bigg( \frac{\delta \rho_{g}}{\delta {\rm t}} \bigg)=
\frac{1}{\rho_{g}}\bigg( -\frac{\delta \rho_{*}}{\delta {\rm t}} \bigg)
= -\frac{\rho_{*}}{\rho_{g}}X_{\rm coll}
\end{equation}

\subsection{Momentum balance}

We modify equation number (2.9) of Langbein et al. (1990) in the following way:
\begin{equation}
\frac{\partial (\rho_{g}u_{g})}{\partial t}=u_{g}\frac{\partial \rho_{g}}{\partial t}+
\rho_{g} \frac{\partial u_{g}}{\partial t};
\end{equation}
\noindent
we substitute this equality in their equation, divide by $\rho_{g}$ ($u_{g}$ 
is the variable in our code) and make use of the equation of continuity for the 
gas component. Thus, we get the following expression:
\begin{equation}
\frac{\partial u_{g}}{\partial t}+u_{g} \frac{\partial u_{g}}{\partial r} + {GM_r\over r^2} +
{1\over\rho_{g}}\frac{\partial p_{r}}{\partial r} -\frac{4 \pi}{c} \kappa_{\rm ext} H= 
\bigg( \frac{\delta u_{g}}{\delta t}\bigg)_{\rm coll}  
\end{equation}
\noindent
To get the interaction term we use the mass and momentum conservation:
\begin{equation}
\bigg( \frac{\delta \rho_{g}}{\delta t}\bigg)_{\rm coll}+
\bigg( \frac{\delta \rho_{*}}{\delta t}\bigg)_{\rm coll}=0 \nonumber
\end{equation}
\begin{equation}
\bigg( \frac{\delta (\rho_{g} u_{g})}{\delta t}\bigg)_{\rm coll}+
\bigg( \frac{\delta (\rho_{*} u_{*})}{\delta t}\bigg)_{\rm coll}=0.
\end{equation}
\noindent
We know that
\begin{equation}
\bigg( \frac{\delta u_{*}}{\delta t}\bigg)_{\rm coll}=0,
\end{equation}
\noindent
thus,
\begin{equation}
\bigg( \frac{\delta (\rho_{g} u_{g})}{\delta t}\bigg)_{\rm coll}=u_{*}\rho_{*}X_{\rm coll}=
\rho_{g} \bigg( \frac{\delta u_{g}}{\delta t}\bigg)_{\rm coll} + u_{g} X_{\rm coll} \rho_{*}.
\end{equation}
\noindent
Therefore, the resulting interaction term is
\begin{equation}
\bigg( \frac{\delta u_{g}}{\delta t}\bigg)_{\rm coll}= \frac{\rho_{*}}{\rho_{g}} 
X_{\rm coll}(u_{*}-u_{g})
\end{equation}
In the case of the stellar system
\begin{equation}
F= \frac{1}{2}(F_{r}+F_{t})=\frac{5}{2}\rho_{*}v_{*}
\end{equation}
By analogy, we now introduce $F_{\rm rad}$ in this way
\begin{equation}
\frac{F_{\rm rad}}{4\pi}=H=\frac{5}{2}p_{g}v_{g},
\end{equation}
where $v_{g}$ is per gas particle.
\begin{equation}
v_{g}=\frac{2}{5} \frac{H}{p_{g}}
\end{equation}

As means to write the equation in its ``logarithmic variable version'', we multiply
the equation by $\rho_{g} r /p_{g}$, as we did for the corresponding momentum balance
star equation and replace $H$ by $5/2p_{g}v_{g}$, 
\begin{eqnarray}
\frac{\rho_{g}r}{p_{g}} \bigg( \frac{\partial u_{g}}{\partial t}+u_{g}
\frac{\partial u_{g}}{\partial r} \bigg)+\frac{GM_{r}}{rp_{g}}\rho_{g}+
\frac{\partial \ln p_{g}}{\partial \ln r}-\frac{5}{2} \frac{\kappa_{\rm ext}}{c}
\rho_{g}r v_{g}= \nonumber \\
\frac{r}{p_{g}} \rho_{*}X_{\rm coll} (u_{*}-u_{g})
\end{eqnarray}

\subsection{Radiation transfer}

We get the radiation transfer equations by re-writing the frequency-integrated moment equations from Amaro-Seoane et al. (2003): We divide
their first equation by $J$ and multiply the second one by $2c/(5p_{g}v_{g})$,
\begin{eqnarray}
\frac{1}{c}\frac{\partial \ln J}{\partial t}+\frac{5}{2J}\frac{\partial}{\partial r} 
(v_{g}p_{g})+\frac{5}{Jr}p_{g}v_{g}-\frac{3f_{\rm Edd}-1}{cr}u_{g} 
-(1+f_{\rm Edd}) \nonumber \\
\frac{1}{c}\frac{\partial \ln \rho_{g}} {\partial t}= 
\frac{\kappa_{\rm abs}}{J} (B-J)
\end{eqnarray}

\begin{eqnarray}
\frac{\partial \ln v_{g}}{\partial t}+\frac{\partial \ln p_{g}}{\partial t}+
\frac{2c}{5}\frac{1}{p_{g}v_{g}}\frac{\partial (J f_{\rm Edd})}{\partial r}+
\frac{2c}{5}\frac{3f_{\rm Edd}-1}{rp_{g}v_{g}}J-\frac{2u_{g}}{r} \nonumber \\
-2\frac{\partial \ln \rho_{g}}{\partial t}= -c \kappa_{\rm ext}\rho_{g} 
\end{eqnarray}
\noindent
Where we have substituted $H=5p_{g}v_{g}/2$ and $\kappa_{\rm abs}$ and $\kappa_{\rm ext}$ are the absorption and extinction
coefficients per unit mass 
\begin{equation}
\kappa_{\rm abs}=\frac{\rho_{g} \Lambda(T)}{B},~
\kappa_{\rm ext}=\rho_{g}(\kappa_{\rm abs}+\kappa_{\rm scatt}),
\end{equation}
\noindent
$\Lambda(T)$ is the cooling function, $B$ the Planck function and $\kappa_{\rm scatt}$ the
scattering coefficient per unit mass. We have made use of $\partial M_{r}/ \partial 
r=4 \pi^2 \rho$, $f_{\rm Edd}=K/J$, 
and the Kirchhoff's law, $B_{\nu}=j_{\nu}/\kappa_{\nu}$ ($j_{\nu}$ is the emission 
coefficient).

\subsection{Thermal energy conservation}

It is enlightening to construct an equation for the energy per 
volume unit $e=(p_{r}+2p_{t})/2$ which, in the case of an isotropic gas ($p_{r}=p_{t}$)
is $e=3p/2$. For this aim we take, for instance, equation (\ref{eqn:pr}) and in the term
$2p_{r}\partial u_{*}/\partial r$ we include now a source for radiation pressure, 
$2(p_{r}+p_{\rm rad})\partial u_{*}/\partial r$ and we divide everything by $e$ so that
we get the logarithmic variables. The resulting equation is

\begin{equation}
\frac{\partial \ln e}{\partial t}+(u_{g}+3v_{g}) \frac{\partial \ln e}{\partial r}
+\frac{2}{r} (u_{g}+v_{g})=\frac{1}{e} \bigg( \frac{\delta e}{\delta t} \bigg)_{\rm drag}+
\frac{1}{e}\bigg( \frac{\delta e}{\delta t} \bigg)_{\rm coll}
\end{equation}

\noindent
The interaction terms for this equation are
\begin{equation}
\bigg( \frac{\delta e}{\delta t} \bigg)_{\rm drag}=X_{\rm drag}(\sigma_{r}^2+\sigma_{t}^2
+(u_{*}-u_{g})^2)
\end{equation}
\begin{equation}
\bigg( \frac{\delta e}{\delta t} \bigg)_{\rm coll}=\frac{1}{2}X_{\rm coll} \rho_{*}
((\sigma_{r}+\sigma_{t})^2 \epsilon+(u_{*}-u_{g})^2 -\xi \sigma_{\rm coll}^2)
\end{equation}
\subsection{Mass conservation}

The mass conservation is guaranteed by

\begin{equation}
\frac{3}{4\pi}\frac{\partial M_{r}}{\partial r^3}=\rho_{*}+\rho_{g}
\end{equation}

\section{A mathematical view of the code}
Our model has seven dependent variables,
\begin{equation}
(\rho, u, v_{r}, v_{t}, p_{r}, p_{t}, M) \equiv {\bf x}(r,t), 
\end{equation}
\noindent
So we have a set of non-linear, coupled differential equations plus an 
initial model, which very often is a Plummer's model. 
To solve our equations we discretise them on a logarithmic radial mesh 
with typically 200 logarithmically equidistant 
mesh points, covering radial scales over eight orders
of magnitude; e.g. from 100 pc down to $10^{-6}$ pc, which is enough to
resolve the system down to the vicinity of a massive black hole's tidal
disruption radius for stars.

An implicit
Newton-Raphson-Henyey iterative method is used to solve for the time
evolution of our system.
Let ${\bf F}(x)$ be a column-vector for the seven equations:
\begin{equation}
\mathbf{F}(x)=
\left( \begin{array}{c}
F_{1}(x_{1} \ldots x_{7}) \\
\vdots \\
F_{7}(x_{1} \ldots x_{7})
\end{array} \right)
\end{equation}
\noindent
The solution to the equations is $x_{\rm true}$; therefore,
\begin{equation}
\mathbf{F}(x_{true})=0.
\end{equation}
\noindent
Suppose that $x^{(1)}$ is a close value
 to the solution $x_{\rm true}$,
\begin{equation}
x^{(1)}=x_{\rm true}+\triangle x;
\end{equation}
\noindent
thus,
\begin{equation}
\mathbf{F}(x^{(1)})=\mathbf{F}(x_{\rm true}+\triangle x)=\mathbf{G}(x),
\end{equation}
\noindent
where $\mathbf{G}(x)$ is an ``error function'' that contains the difference
between the exact value and the approximation.
Since we have assumed that $x^{(1)}$ is a close value to $x_{\rm true}$,
\begin{equation}
\mathbf{F}(x_{\rm true}+\triangle x) \approx \mathbf{F}(x_{\rm true})+
\frac{\partial \mathbf{F}}{\partial x} \triangle x=\mathbf{G}(x),
\end{equation}
\noindent
therefore, since $\mathbf{F}(x_{\rm true})=0$,
\begin{equation}
\triangle x=\mathbf{G}(x) \bigg( \frac{\partial \mathbf{F}}{\partial x}  \bigg)^{-1},
\end{equation}
\noindent
where
\begin{equation}
\frac{\partial \mathbf{F}}{\partial x} \equiv 
\left( \begin{array}{ccc}
{\partial F_{1}(x_{1} \ldots x_{7})}/{\partial x_{1}} 
& & \\
& \ddots & \\
& & {\partial F_{7}(x_{1} \ldots x_{7})}/{\partial x_{7}}
\end{array} \right)
\end{equation}
\noindent
is a $7 \times 7$ matrix. We iterate the process until we reach
\begin{equation}
\bigg| \frac{\triangle x^{(n)}}{x^{(n)}} \bigg| \le \varepsilon \sim 10^{-6},
\end{equation}
where $n$ stands for the n-iteration done. This gives us the termination of the
iteration.
Our results compare well with other studies
using direct solutions of the Fokker-Planck equation or Monte Carlo models
(Lightman \& Shapiro 1977, Marchant \& Shapiro 1980). Note that the
Monte Carlo approach has been recently revisited and improved by
Freitag \& Benz (2001). In contrast to the other models the gaseous
model is much more versatile to include all kinds of important other
physical effects, such as the dynamics of gas liberated in nuclei
by stellar evolution and collisions and its interaction with the
stellar component.

\def\aa{{A\&A}}
\def\aas{{A\&AS}}
\def\aj{{AJ}}
\def\annrev{{ARA\&A}}
\def\apj{{ApJ}}
\def\apjs{{ApJS}}
\def\baas{{BAAS}}
\def\mnras{{MNRAS}}
\def\nat{{Nature}}
\def\pasp{{PASP}}

\end{document}